\pdfoutput=1
\documentclass[journal,twoside,web]{ieeecolor}
\let\labelindent\relax
\usepackage{generic}
\usepackage{cite}
\usepackage{amsmath,amssymb,amsfonts}
\usepackage{graphicx}
\usepackage{textcomp}
\usepackage{booktabs}
\usepackage{tabularx}
\usepackage{url}

\usepackage[ruled,vlined]{algorithm2e}
\usepackage{algpseudocode}
\usepackage{bm}
\usepackage{enumitem}
\setlist[itemize]{leftmargin=*}

\def\BibTeX{{\rm B\kern-.05em{\sc i\kern-.025em b}\kern-.08em
    T\kern-.1667em\lower.7ex\hbox{E}\kern-.125emX}}
\begin{document}

\title{Joint Task Offloading and Channel Allocation in Spatial-Temporal Dynamic for MEC Networks}
\author{
Tianyi Shi, \IEEEmembership{Graduate Student Member, IEEE}, 
Tiankui Zhang, \IEEEmembership{Senior Member, IEEE},
Jonathan Loo,
Rong Huang,
Yapeng Wang, \IEEEmembership{Member, IEEE}
\thanks{This work was supported by the National Natural Science Foundation of China (No. 62371068).}
\thanks{Tianyi Shi and Tiankui Zhang are with the School of Information and Communication Engineering, Beijing University of Posts and Telecommunications, Beijing 100876, China (e-mail: shi\_tianyi@bupt.edu.cn; zhangtiankui@bupt.edu.cn).}
\thanks{Jonathan Loo is with the School of Electronic Engineering and Computer Science, Queen Mary University of London, London E1 4NS, U.K., (e-mail:j.loo@qmul.ac.uk).}
\thanks{Rong Huang is with the China Unicom Research Institute, Beijing 100089, China (e-mail: huangr27@chinaunicom.cn).}
\thanks{Yapeng Wang is with the Faculty of Applied Sciences, Macau Polytechnic University, Macau 999078, China (e-mail: yapengwang@mpu.edu.mo).}
}
\maketitle

\begin{abstract}
Computation offloading and resource allocation are critical in mobile edge computing (MEC) systems to handle the massive and complex requirements of applications restricted by limited resources. 
In a multi-user multi-server MEC network, the mobility of terminals causes computing requests to be dynamically distributed in space. At the same time, the non-negligible dependencies among tasks in some specific applications impose temporal correlation constraints on the solution as well, leading the time-adjacent tasks to experience varying resource availability and competition from parallel counterparts.
To address such dynamic spatial-temporal characteristics as a challenge in the allocation of communication and computation resources, we formulate a long-term delay-energy trade-off cost minimization problem in the view of jointly optimizing task offloading and resource allocation. 
We begin by designing a priority evaluation scheme to decouple task dependencies and then develop a grouped Knapsack problem for channel allocation considering the current data load and channel status.
Afterward, in order to meet the rapid response needs of MEC systems, we exploit the double duel deep Q network (D3QN) to make offloading decisions and integrate channel allocation results into the reward as part of the dynamic environment feedback in D3QN, constituting the joint optimization of task offloading and channel allocation. Finally, comprehensive simulations demonstrate the performance of the proposed algorithm in the delay-energy trade-off cost and its adaptability for various applications.

\end{abstract}


\section{Introduction}
\label{sec:introduction}
\IEEEPARstart{B}{y} deploying the computing and storage resources at the network's edge, which is closer to the terminals, mobile edge computing (MEC) satisfies the demanding performance requirements of various emerging delay-sensitive applications such as virtual reality, industrial detection, e-medicine, and autonomous driving \cite{1,ref2025_1}. However, the related resource at the edge is far less adequate than that of the cloud server \cite{2}, so it is essential to design reasonable mechanisms when offloading computational tasks to suffice the business of mobile applications and guarantee the quality of service (QoS). Moreover, it avoids excessive network overhead and loads while improving resource utilization. 

The present research on task offloading in MEC networks mainly focuses on two aspects: offloading strategy and resource management. 
The main goal of the former is to optimize the system delay, energy consumption, and the balance of the load. 
Utilizing various algorithms such as convex optimization, Lyapunov optimization, heuristics, and the greedy method~\cite{202401,202402,2024comparison,2025_heuristic,9257019}, the system efficiently resolves user decisions regarding whether and what tasks to offload.
How to configure the limited resources is also crucial to the performance of the MEC systems. 
Some studies~\cite{8877759,2024channel_knapsack,202403} also optimize resource allocation from the aspects of node assignment and deployment, network status awareness, and architecture design to improve the overall network performance.
Some works focus on load balancing of the cloud-fog-edge system and adopt meta-heuristics and hybrid methods to make full use of system resources and improve the overall QoS ~\cite{reviewer402,reviewer401,reviewer403}.
However, with the refinement of research, new difficulties have come up with this issue in both space and time scales.

Regarding the spatial dynamics of the environment, actual MEC networks frequently experience high levels of dynamism because of wireless channel fluctuations and terminal mobility. 
Moreover, the movement of terminals exerts an impact on the traffic load of each MEC node, so the radio resource allocation, as a key factor of the system capability, becomes more crucial and intractable. 
In the case that terminals can only grasp local information, traditional solutions cannot respond quickly or make decisions in real-time for the MEC system~\cite{ref2025_1}. 
The appearance of deep reinforcement learning~(DRL) alleviates the above dilemma and provides a solution to optimize long-term rewards for computation offloading in dynamic scenarios~\cite{20,9434397,spatial1,9916157}.
The authors of~\cite{spatial1} proposed a partial offloading algorithm for delay-sensitive tasks on the basis of Q-learning to make discrete offloading decisions and deep deterministic policy gradient~(DDPG) for continuous offloading decisions to provide more flexibility.
The authors of~\cite{9916157} investigated the joint problem of collaborative task offloading and resource allocation between the MEC and the cloud servers and proposed a deep Q network (DQN)-based algorithm that combines double DQN, duel DQN, and adaptive parameter space noise to optimize the total energy consumption. 
Notably, associating double and duel DQN~(D3QN)~\cite{19} possesses both the stability of double DQN and the efficiency of duel DQN, which is a good option to meet the demands for real-time decisions of MEC systems in the intricate environment.
As we reviewed, most existing works often neglected the spatial mobility of terminals and lacked active resource management strategies, which are essential for ensuring flexible and efficient utilization of the limited resources within the MEC system.
\par

From the perspective of time correlation, it is essential to acknowledge the inherent execution sequence among different modules within the application when delivering increasingly advanced mobile services.
Taking autonomous driving as an example, it can generally be divided into four modules: perception, positioning, planning while decision-making, and control. In such cases, the application should be considered as dependently seperated tasks rather than a single entity and scheduled based on their chronological sequences.
To deal with this issue, most existing works model tasks with dependencies as a directed acyclic graph (DAG) and decouple them using priority-based methods ~\cite{DAGnew,new1,new2}. 
In \cite{DAGnew}, the authors introduced a DRL-based task scheduling algorithm in a vehicular edge computing network to minimize the long-term system delay and energy consumption cost. 
The authors of \cite{new1} proposed an actor-critic-based computational offloading scheme for dependent IoT applications with a prioritized scheduling strategy to achieve low latency.
Leveraging specialized neural networks, the authors of \cite{14new} adopted a sequence-to-sequence neural network to extract features of task dependencies and proposed a DRL-based offloading framework to reduce the latency and energy consumption, while the authors of \cite{15} employed a graph convolutional neural network to capture the task structures in a multi-user MEC scenario with wireless interference.
Our observations indicate a research gap in offloading strategies for tasks with dependencies, particularly within a comprehensive network architecture that considers dynamic competition and collaboration over communication and computing resources among multiple entities.

\par

Although existing studies have considered the spatial dynamics and temporal correlation of tasks separately, they have not addressed both aspects together or incorporated active resource management.
To fully utilize communication and computing resources in MEC networks and enhance performance in dynamic environments, cohesive strategies that adapt to the mobility of terminals and the random fading of the channel for task offloading and resource allocation are essential.
To the best of our knowledge, this work is the first to comprehensively consider these challenges and provide a learning-based solution capable of real-time adaptation, tackling topological dynamics in both the offloading request distribution and the resource availability.
The specific contributions of our work are as follows.

\begin{itemize}
    \item An MEC architecture with spatial-temporal characteristics is defined, in which multiple edge servers offer communication and computing resources for multiple mobile terminals. We consider the co-channel interference among cells and the parallel computing with I/O interference of the edge server in the model. A priority estimation rule according to average cost is designed to transform the tasks in DAGs into topological sequences, which guarantees the dependencies among tasks. Then, we formulate a long-term delay-energy trade-off cost minimization problem by jointly optimizing offloading decisions and channel allocation.
    \item To deal with the above problem, we propose a D3QN-based task offloading algorithm integrated multi-cell channel allocation (TOICA) by solving the two variables separately. First, a dynamic programming-based channel allocation algorithm (DCA) is put forward to optimize the system resource configuration, which is modeled into a grouped Knapsack problem. Then, for real-time decision and long-term optimization, we develop a D3QN-based task offloading algorithm (DTO). Lastly, the coupling details of the two and the implementation of the whole proposed TOICA algorithm are elaborated.
    \item Simulations are provided to validate the performance of the proposed TOICA compared with other benchmarks. The results also verify the effectiveness of the proposed algorithms under different relevant parameters on the system cost reduction and compare the trade-offs between latency and energy weights for various applying purposes.
\end{itemize}

The organization of this work is as follows. The system model is described in section \uppercase\expandafter{\romannumeral2}. Section \uppercase\expandafter{\romannumeral3} formulates the optimization problem. The channel allocation and task offloading algorithms are proposed in section \uppercase\expandafter{\romannumeral4} and section \uppercase\expandafter{\romannumeral5} illustrates the simulation results.
\section{System Model and Problem Formulation}
This section presents the models involved and corresponding formulations in this work and the notations are summarized in Table \ref{table0}.

\begin{table}[t]
	\caption{Notations}
        \begin{tabular}{m{2.2cm}|m{5.8cm}}
            \toprule[1pt]
             \bf{Notations}&\bf{Definitions}  \\
             \hline
             $\mathcal{N}, n, N$&Set, index and total number of MTs \\
             $\mathcal{M}, m, M$&Set, index and total number of ESs \\
             $\mathcal{K}, k, K$&Set, index and total number of subchannels\\
             
             $G_n=(V_n,E_n)$&The DAG of mobile application on MT $n$, the set of tasks and the set of temporal relationships \\
             $I_n,i$&The total number of divided tasks of MT $n$ and task's index \\
             $v_{n,i}, e_{i,j}$&Task $i$ of MT $n$, the existence of the relationship between $v_{n,i}$ and $v_{n,j}$\\
             
             $\mathcal{T}, t, T, D[t]$&Set, index, the total number of time slots, and its duration, ($T=\max\limits_{n}{I_n}$)\\
             
             $\textbf{entry}({G_n})$& Set of entry nodes of $G_n$\\
             $\textbf{exit}({G_n})$& Set of exit nodes of $G_n$\\

             $v_{n,p}, p\in\textbf{pre}(i)$ & The immediate predecessor of $v_{n,i}$, the set of $p$\\
             $v_{n,q}, q\in\textbf{succ}(i)$ & The immediate successor of $v_{n,i}$, the set of $q$\\
             
             $b_{n,i}, c_{n,i}$ & The input data volume and the required computation volume of task $v_{n,i}$\\
             $RT_{n,i}, FT_{n,i}$ & Ready time, finish time of task $v_{n,i}$\\
             $o_{n,i,m}[t]$, $x_{n,k}[t]$ &  Indicator whether task $v_{n,i}$ is offloaded to ES $m$ at slot $t$, indicator whether subchannel $k$ is assigned to MT $n$ at slot $t$ \\
             $N_m[t], \mathcal{N}_m[t]$ & The total number and the set of MTs that decided to offload tasks to ES $m$ at slot $t$ \\
             $d_{n,i,0}, e_{n,i,0}$ & Completion time and energy consumption of task $v_{n,i}$ processed locally\\
             $d_{n,i,m}, e_{n,i,m}$ & Completion time and energy consumption of task $v_{n,i}$ processed by ES $m$\\
             $f_n,f_m$ & Computation capacity of MT $n$ and ES $m$\\
             $p_{n,k}^{tr}, p_{n}^{tr}, p_{n}^{st}$ & Transmission power on subchannel $k$, total transmission power and the static power of MT $n$\\
             $cost_{n,i}$ & Cost of task $v_{n,i}$\\
             $\mathcal{P}_{n,i}$ & Priority of task $v_{n,i}$ \\
             \bottomrule[1pt]
        \end{tabular}
        \label{table0}
\end{table}

\subsection{Scenario Description}
\begin{figure} [t]
\centerline{\includegraphics[width=\columnwidth]{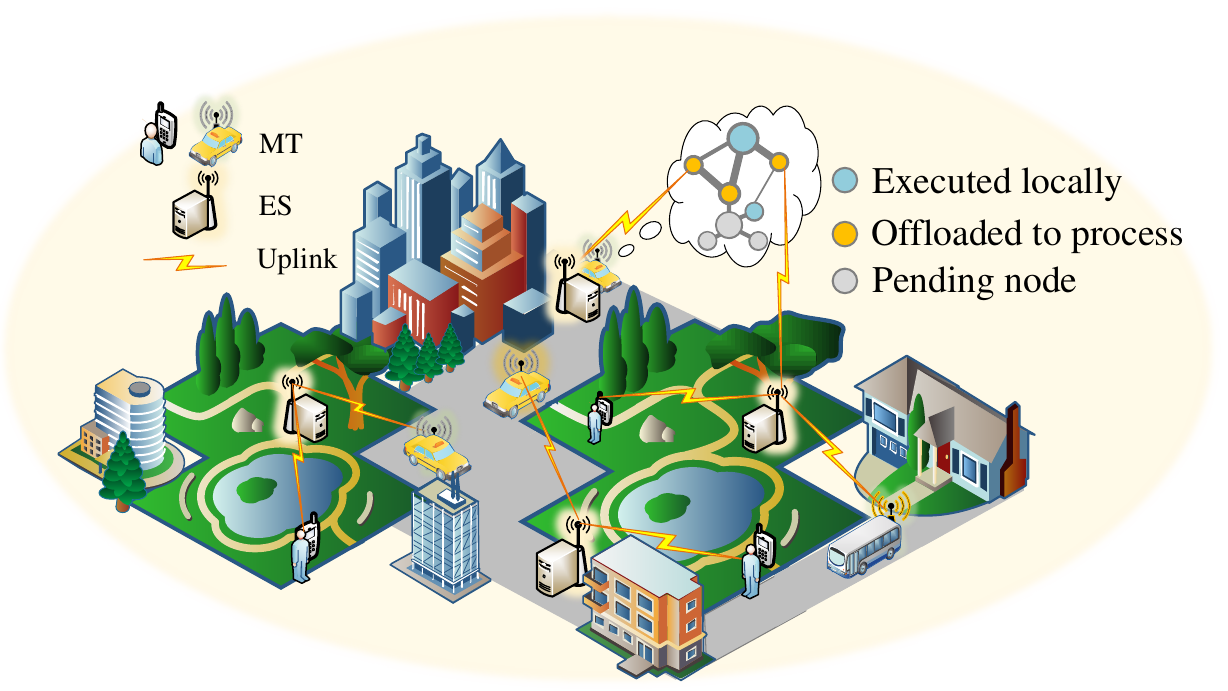}}
\caption{Spatial and temporal dynamic MEC networks.}
\label{fig1systemmodel}
\end{figure}

As shown in Fig.~\ref{fig1systemmodel}, there are $N$ mobile terminals (MT) denoted as $\mathcal{N}=\left\{1,2,\ldots,n,\ldots,N\right\}$, which may be pedestrians' smartphones or smart vehicles on the road. These MTs move randomly, implying that their positions, denoted by $\textbf{X}_{n}\in\mathbb{R}^2, (\forall{n}\in\mathcal{N})$, are not predetermined. $M$ edge servers (ES) denoted by $\mathcal{M}=\left\{1,2,\ldots,m,\ldots,M\right\}$, with fixed positions represented by $\textbf{Y}_n\in\mathbb{R}^2, (\forall{m}\in\mathcal{M})$, that can communicate with accessed MTs. $K$ orthogonal subchannels are divided for each cell to transmit data between MTs and ESs, while every subchannel's bandwidth is $B_k$. 
\par

\begin{figure} [t]
\centerline{\includegraphics[width=\columnwidth]{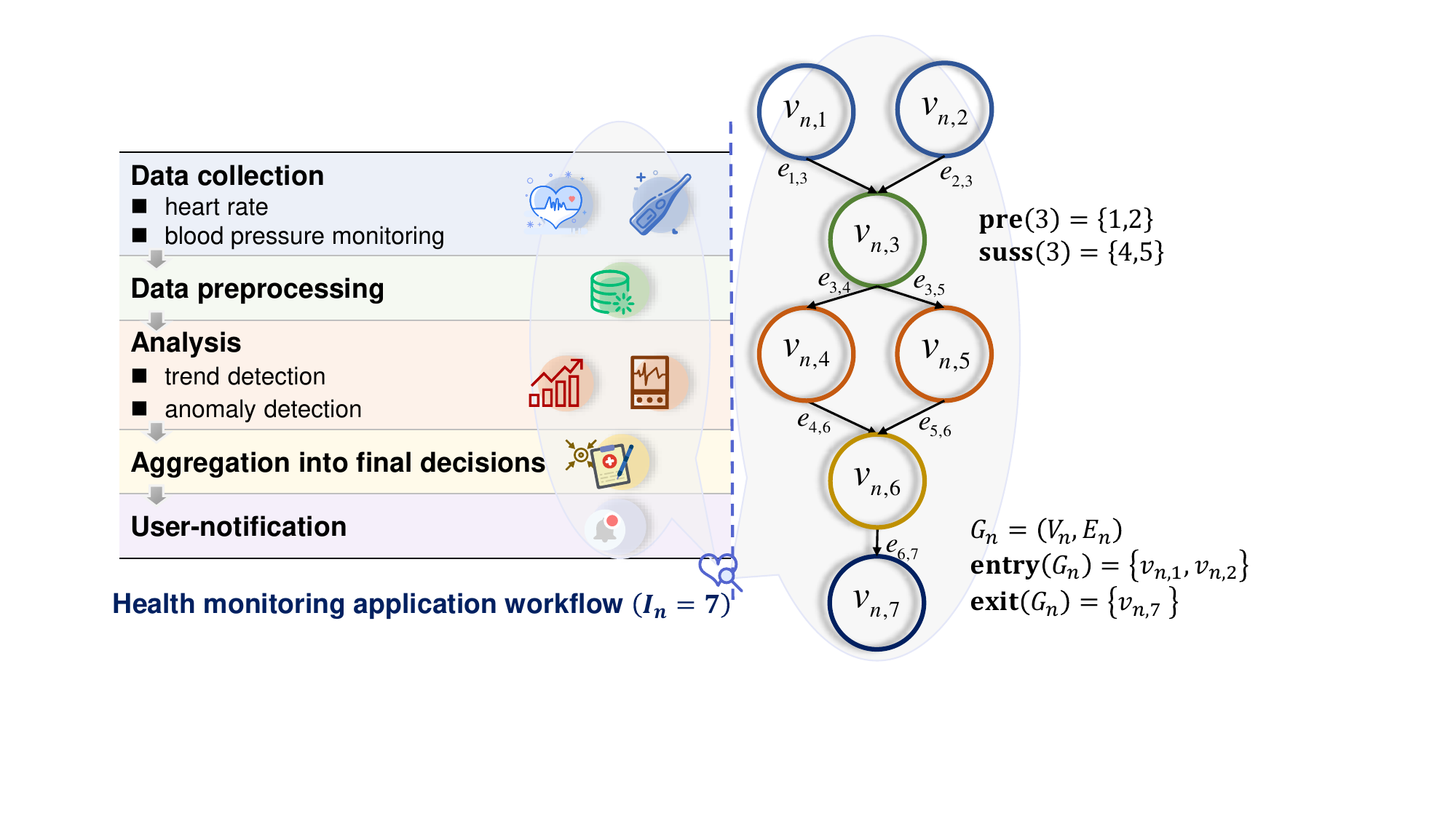}}
\caption{DAG structure of task dependencies (Health monitoring application example).}
\label{fig2DAGstructure}
\end{figure}

A mobile application on MT $n$ with computing demand is partitioned into $I_n$ tasks, which can be executed locally on terminals or offloaded to an ES for processing. 
These $I_n$ tasks on MT $n$ are represented by $V_n=\left\{v_{n,i}| \forall{i}\in\mathcal{I}_n\right\}$, with dependencies among them are denoted as $E_n$.
Specifically, $e_{i,j}\in{E_n}$ indicates that task $v_{n,j}$ can only be executed after task $v_{n,i}$ has been processed.
Thus, these tasks and their interdependencies can be represented by a directed acyclic graph $G_n=(V_n, E_n)$, as illustrated in Fig.~\ref{fig2DAGstructure}. 
The task dependencies in the DAG are application-specific and can be obtained using standard profiling techniques, such as performance monitoring tools (e.g., Android Profiler, Xcode Instruments) or manual identification.
For example, a health monitoring application consists of several stages, including data collection (e.g., heart rate and blood pressure), preprocessing, analysis (such as trend and anomaly detection), result aggregation, and final user notification, and these tasks are organized with specific dependencies.
The entry task set $\textbf{entry}({G_n})$ in the DAG marks the start of the workflow, with no preceding tasks ($e_{i,en} \notin E_n$), while the exit task set $\textbf{exit}(G_n)$ represents the final stage with no succeeding tasks ($e_{ex,i} \notin E_n$).
If $e_{p,i}\in E_n$ for all ${p}\in\mathcal{I}_n$, then $v_{n,p}$ is the immediate predecessor of task $v_{n,i}$, and we define $p\in\textbf{pre}(i)$. Similatrly, if $e_{i,q}\in E_n$ for all ${q}\in\mathcal{I}_n$, then $v_{n,q}$ is the immediate successor of task $v_{n,i}$, and we define $q\in\textbf{succ}(i)$. For instance, the predecessors of $v_{n,3}$ are $v_{n,1}$ and $v_{n,2}$, while its successors are $v_{n,5}$ and $v_{n,6}$. 
In this context, the output of data collection serves as input for data preprocessing, and the results of tasks like trend and anomaly detection are combined to generate the final user notification.
The computational task $v_{n,i}$ can be described as a 2-tuple ($b_{n,i}$, $c_{n,i}$), in which elements represent the input data volume (in bits) and the required computation volume (in CPU cycles), respectively.
It's important to note that the specific DAG structure, tailored to the application's functionality and workflow, does not impact the efficacy of the proposed underlying models and methodologies.
Additionally, we assume that each task is completed within a single time slot, and the duration of slot $t$ is denoted as $D[t], t\in\mathcal{T}=\left\{1,2,\ldots,t,\ldots,T\right\}$, with $T=\mathop{\max}\limits_{n}{I_n}$.

\subsection{Task Dependency Model}
According to the task nodes' structure of DAG, task $v_{n,i}$ cannot be executed until all of its immediate predecessors have been processed. The ready time $RT_{n,i}$ of $v_{n,i}$ is expressed as
\begin{equation}
    RT_{n,i}=\mathop{\max}_{p\in\textbf{pre}(i)} FT_{n,p}.\label{(1)}
\end{equation}
It indicates from the perspective of DAG structure the moment that $v_{n,i}$ can be processed, which is the maximum finish time among all predecessors of $v_{n,i}$. Explicitly the finish time $FT_{n,i}$ of task $v_{n,i}$ is derived as
\begin{equation}
    FT_{n,i}=RT_{n,i}+d_{n,i,m^{\prime}},\label{(2)}
\end{equation}
where $m^{\prime}\in\{0\}\cup\mathcal{M}$ means a certain way of handling the task: MT $n$ executes it locally when $m^{\prime}=0$, and $m^{\prime}=m$ implies the task is offloaded to ES $m$ for processing.

The completion sign of mobile application $G_n$ is all of its exit tasks have been processed. Thus, its execution delay $delay_n$ can be expressed as
\begin{equation}
    delay_n=\mathop{\max}_{v_{n,ex}\in\textbf{exit}(G_n)} FT_{n,ex}.\label{(3)}
\end{equation}

\subsection{Task Processing Model}
We define $o_{n,i,m}[t]\in\{0,1\}$ as an offloading indicator for task $v_{n,i}$. Here, $o_{n,i,m}[t]=1$ suggests that task $v_{n,i}$ is offloaded to ES $m$ to be processed at slot $t$, while $\sum_{m}^{M}o_{n,i,m}[t]=0$ means that it is executed locally. 
There is another variable, $x_{n,k}[t]\in\{0,1\}$, is used to define the allocation of subchannels. When $x_{n,k}[t]=1$, it represents that subchannel $k$ is assigned to MT $n$ at slot $t$; otherwise, it isn't occupied by the terminal.

\subsubsection{Processing locally}
If the computational task $v_{n,i}$ is processed locally, the task completion time is only decided by the computing time of MT $n$, which is
\begin{equation}
    d_{n,i,0}=\frac{c_{n,i}\sum_t^T \left\{1-\sum_{m=1}^M o_{n,i,m}[t]\right\}}{f_n},\label{(4)}
\end{equation}
and the corresponding energy consumption is
\begin{equation}
    e_{n,i,0}=\kappa c_{n,i}\sum_{t=1}^T \left\{1-\sum_{m=1}^Mo_{n,i,m}[t]\right\} {f_n}^2,\label{(5)}
\end{equation}
where $f_n$ is expressed as the computing power (CPU frequency) of MT $n$. $\kappa$ is the switched capacitance.

\subsubsection{Processing at edge servers}
If task $v_{n,i}$ is offloaded to ES $m$, both the communication procedure that MT $n$ transmits data $b_{n,i}$ and the computation procedure of the server need to be taken into consideration. 
Technically, communication time $d_{n,i}^{tr}$ encompasses both the uploading of input data and the returning of outputs. 
However, in alignment with the rationale adopted in related works~\cite{2024comparison} and \cite{2024ignore_downlink}, our model assumes negligible downlink latency. This assumption is grounded on two key observations. First, the output data generated by mobile applications is typically much smaller than the uploaded input data. Second, the advancements in modern cellular networks, with their high reliability and fast downlink speeds, further justify this assumption, rendering downlink latency negligible to overall system performance. Thus, we disregard downlink latency and focus on the more critical aspects of uplink transmission and computation times.

The channel gain between terminals and edge servers is determined by large-scale path loss and small-scale fading due to the movement of MTs. In slot $t$, the uplink data rate between MT $n$ and ES $m$ is
\begin{equation}
    r_{n,m}^{up}[t]=\sum_{k=1}^K B_{k}\log_2 \left(1+\frac{x_{n,k}[t]p_{n,k}^{tr}{h_{n,m,k}}[t]}{I_{n,k}[t]+\sigma^2}\right),\label{(6)}
\end{equation}
where $I_{n,k}[t]=\sum\limits_{\widetilde{m}\in{\left\{\mathcal{M}\backslash{m}\right\}}}\sum\limits_{\widetilde{n}\in{\mathcal{N}_{\widetilde{m}}}} x_{\widetilde{n}},k[t]p_{\widetilde{n}},k^{tr}\beta_{k}[t]l_{\widetilde{n}},m[t]^{-\alpha}$ and it represents co-channel interference caused by other MTs $\widetilde{n}$ that utilize the same subchannel $k$, while $\mathcal{N}_{\widetilde{m}}$ is the set of MTs that access other ES, denoted as $\widetilde{m}$, distinct from the target ES $m$.
The term $p_{n,k}^{tr}$ represents the transmission power of MT $n$ on subchannel $k$, distributed evenly among its total transmission power.
$\beta_{k}[t]$ is an independent random variable with an exponential distribution with a unit mean, representing the Rayleigh fading of channel $k$. $l_{n,m}[t]=\lvert\textbf{Y}_m-\textbf{X}_{n}[t]\rvert$ equals to the current Euclidean distance between MT $n$ and ES $m$, and $\alpha$ is the path loss index. Then, the channel gain $h_{n,m,k}[t]$ between MT $n$ and ES $m$ on channel $k$ in slot $t$ is defined as $h_{n,m,k}[t]=\beta_{k}[t]l_{n,m}[t]^{-\alpha}$. $\sigma^2$ refers to the additive Gaussian noise power.

Therefore, the communication time of task $v_{n,i}$ of MT $n$ that offloaded to ES $m$ for processing is
\begin{equation}
    d_{n,i,m}^{tr}=\sum_{t=1}^T \frac{b_{n,i}o_{n,i,m}[t]}{r_{n,m}^{up}[t]}.\label{(7)}
\end{equation}

ES $m$ performs parallel computing on tasks from different terminals in its cache by dynamically creating and deleting a relevant number of virtual machines (VM). Considering that the I/O interference among the concurrent VMs will affect the processing speed of tasks \cite{16}, the computing time of $v_{n,i}$ processed by edge server $m$ can be expressed as
\begin{equation}
    d_{n,i,m}^{co}=\sum_{t=1}^T D_{0}o_{n,i,m}[t](1+\zeta)^{N_m[t]-1},\label{(8)}
\end{equation}
in which $N_m[t]$ means the current number of terminals accessing to ES $m$, $D_0=c_{n,i}/f_m$ represents the required processing time of ES $m$ with one VM ($N_m[t]=1$). $f_m$ means the computing capacity of ES $m$ and $\zeta$ is the degradation factor resulting in the I/O interference among VMs.

Thus, the total time of processing task $v_{n,i}$ by ES $m$ is expressed as the sum of communication and computation time as follows

\begin{equation}
\begin{aligned}
           d_{n,i,m}&=d_{n,i,m}^{tr}+d_{n,i,m}^{co}.\label{(9)}
\end{aligned}
\end{equation}
The corresponding energy consumption of MT $n$ is
\begin{equation}
    e_{n,i,m}=p_{n}^{tr}d_{n,i,m}^{tr}+p_n^{st}d_{n,i,m}^{co},\label{(10)}
\end{equation}
where $p_{n}^{tr}$ and $p_n^{st}$ are the total transmission power and the static power of MT $n$, respectively.

According to the task processing models constructed above, the actual delay-energy trade-off cost of $v_{n,i}$ is
\begin{equation}
    cost_{n,i}=\omega\sum_{m^{\prime}=0}^{M} d_{n,i,{m^{\prime}}}+(1-\omega)\sum_{m^{\prime}=0}^M e_{n,i,{m^{\prime}}},\label{(11)}
\end{equation}
where $\omega$ is a trade-off coefficient between delay and energy and ${d}_{n,i,m^{\prime}}$ and ${e}_{n,i,m^{\prime}}$ serve as the delay and energy consumption, respectively, for processing $v_{n,i}$ in mode $m^{\prime} \left(\forall m^{\prime} \in  \{0,1,\ldots,M\}\right )$.

\subsection{Task Priority}
Echoing the prioritization approach in \cite{DAGnew}, we design a rule for estimating tasks' priority resting on the maximum expected cost. First, the average cost of processing task $v_{n,i}$ is calculated
\begin{equation}
    \overline{COST}_{n,i}=\sum_{m^{\prime}=0}^M {cost}_{n,i,m^{\prime}},\label{(12)}
\end{equation}
where ${cost}_{n,i,m^{\prime}}$ serves as the estimated delay-energy balanced cost of processing $v_{n,i}$ in the way of $m^{\prime}$, with ${cost}_{n,i,m^{\prime}} = \omega d_{n,i,{m^{\prime}}} + \left(1-\omega\right) e_{n,i,{m^{\prime}}}$.
Note that when evaluating the average cost, the access relationship of subchannels between other terminals and servers isn't taken into account while the co-channel interference from adjacent cells is also temporarily ignored. On this basis, the priority $\mathcal{P}_{n,i}$ of $v_{n,i}$ is defined as
\begin{equation}
    \mathcal{P}_{n,i}=\left\{
    \begin{array}{rcl}
    \overline{COST}_{n,i},\quad{v_{n,i}\in{\textbf{exit}(G_n)}}\\
    \underset{{s\in{\textbf{succ}(i)}}}{{\max}}\mathcal{P}_{n,s}+\overline{COST}_{n,i},\quad{v_{n,i}\notin{\textbf{exit}(G_n)}}.\label{(13)}
    \end{array}
    \right.
\end{equation}
Specifically, when $v_{n,i}$ is an exit task, $\overline{COST}_{n,i}$ represents its priority level directly; otherwise, it is the sum of the highest priority of its immediate successors and $\overline{COST}_{n,i}$. Hence, the priority of all tasks can be obtained by recursively calculating from the exit task. Through ranking $\mathcal{P}_{n,i}$ in descending order, the dependencies of tasks of the application $G_n$ could be guaranteed by scheduling the topological queue sequentially.

\section{Task Offloading and Channel Allocation Jointly Optimization }
To satisfy the computing demand of delay-sensitive applications while reducing the energy consumption of terminal devices, we formulate the joint optimization of computational task offloading and subchannel allocation with spatial-temporal task dependencies as problem \eqref{(14)}.
\begin{align}
    \underset{\bf{o},\bf{x}}{\min}\quad&\omega\sum_{t}^{T} D[t]+\sum_{n}^{N} \sum_{i}^{I_n} (1-\omega)(e_{n,i,0}+e_{n,i,m}),\label{(14)}\\
    \text{s.t.}&\nonumber\\
    &D[t]=\max\left\{d_{n,i,0},d_{n,i,m}\big|o_{n,i,m}[t]=1\right\},&\nonumber\\
    &\qquad\qquad\qquad\qquad\qquad\qquad\forall{n,m,i},\tag{14a}\\
    &o_{n,i,m}[t]\in\left\{0,1\right\},\forall{n,m,i},\tag{14b}\\
    &\sum\nolimits_{m}^{M}o_{n,i,m}[t]\leq1,\forall{n,i},\tag{14c}\\
    &x_{n,k}[t]\in\left\{0,1\right\},\forall{n,k},\tag{14d}\\
    &\sum\nolimits_{n\in{\mathcal{N}_m}}x_{n,k}[t]\leq1,\forall{m,k},\tag{14e}\\
    &\sum\nolimits_{n\in{\mathcal{N}_m}}\sum\nolimits_{k=1}^K x_{n,k}[t]=K,\forall{m},\tag{14f}\\
    &\sum\nolimits_{k=1}^K x_{n,k}[t]\geq1,\forall{m,n\in{\mathcal{N}_m[t]}},\tag{14g}\\
    &RT_{n,i}\geq FT_{n,p},\forall{i,p\in{\textbf{pre}(i)}}.\tag{14h}
\end{align}
In \eqref{(14)}, constraint (14a) imposes a limit on the duration of each time slot, guaranteeing that every terminal’s operation (whether local or offloaded) completes within a single slot. Constraint (14b) denotes the binary offloading decision for MT $n$ at slot $t$ to ES $m$, while (14c) ensures that each terminal can offload tasks to at most one server in each slot. Constraints (14d) and (14e) define the binary attributes of channel allocation, ensuring each subchannel can be assigned to at most one MT. Concerning the full utilization of channels, (14f) requires that all $K$ subchannels be occupied by ES $m$, and (14g) ensures that every terminal deciding to offload receives at least one subchannel for data transmission. Finally, (14h) captures the dependency among tasks, specifying that the ready time of $v_{n,i}$ cannot precede the finish time of any of its immediate predecessors.

\par

The joint optimization problem in (14) involves binary decision variables for both task offloading and channel allocation, along with constraints that capture spatial-temporal task dependencies and limited resource availability, naturally exhibiting the combinatorial complexity typical of NP-hard problems.
Its NP-hardness is established by demonstrating that it is reducible from a known NP-hard problem, as elaborated in Section~\uppercase\expandafter{\romannumeral3}-A.
However, practical MEC networks demand rapid responses to accommodate advanced applications, even in the face of such complexity. To address this challenge, we propose a holistic approach that optimizes both channel allocation and task offloading in a coupled manner, ensuring efficient resource utilization within the ever-evolving MEC environment.

\subsection{Dynamic Programming-based Channel Allocation}\label{solve_channel_allocation}
Given that offloading decisions $\mathbf{o}$ are fixed, ES $m$ must judiciously allocate its $K$ subchannels among the $N_m$ MTs that have offloaded tasks to it. Since the amount of data to be processed is determined by the offloading decisions, our objective in this subproblem shifts towards minimizing the corresponding communication cost, which depends on how the subchannels are configured and aligns with the overall delay-energy trade-off goal.
Inspired by \cite{2024channel_knapsack}, we reformulate the subchannel allocation problem into a form reducible to the \emph{grouped knapsack problem} (GKP), a classical NP-hard problem, allowing us to adopt a pseudo-polynomial-time dynamic programming approach. Evolving from the fundamental 0-1 knapsack, GKP partitions items into groups with the constraint that at most one item from each group can be selected. This mechanism is well-suited for allocating the number of subchannels to each terminal while satisfying constraint (14g).
\par

\emph{GKP formulation for subchannel allocation:}
Each \emph{group} is equivalent to a terminal $n \in \mathcal{N}_m$. Let $\mathcal{Z} = \{1,2,\dots,K - N_m+1\}$ represent the potential number of subchannels (i.e., \emph{items}) assignable to a single terminal. We define a binary variable $y_{n,z} \in \{0,1\}$ that indicates whether terminal $n$ is assigned $z$ subchannels. Let $v_{n,z}$ be the resulting communication cost (i.e., the \emph{value} of the item), while $z$ itself corresponds to the \emph{weight} in knapsack terminology. We can then formulate the GKP subproblem as

\begin{align}
    \min\limits_{\{y_{n,z}\}}&\sum\limits_{n\in\mathcal{N}_m}\sum\limits_{z\in\mathcal{Z}_n} v_{n,z}y_{n,z},\label{eq_sp1}\\
    \text{s.t. }& \sum\limits_{z\in\mathcal{Z}_n}y_{n,z}=1, \forall{n},\nonumber\tag{15a}\\
    \quad\quad& \sum\limits_{n\in\mathcal{N}_m}\sum\limits_{z\in\mathcal{Z}_n}z y_{n,z}=K.\nonumber\tag{15b}
\end{align}

In alignment with (14g), constraint (15a) ensures exactly one item is chosen per group, forcing each offloaded terminal to be allocated at least one subchannel. Constraint (15b), echoing (14e) and (14f), guarantees that all $K$ subchannels are fully utilized without exceeding the system’s capacity.
\par
Moreover, since GKP is a well-known NP-hard problem~\cite{kellerer2004knapsack}, and our channel allocation subproblem can be directly mapped to a GKP instance in polynomial time, it follows that the overall problem~\eqref{(14)} also inherits NP-hardness.
In this manner, discrepancies in available bandwidth and channel conditions can be accommodated for diverse resource requirements, while the final allocation also provides a basis for co-channel interference considerations when performing environment interactions, as illustrated in~Section~\uppercase\expandafter{\romannumeral3}-B.
Dynamic programming is a widely used approach for solving the knapsack problem efficiently by systematically exploring possible combinations to find the optimal solution. It iteratively updates a value matrix to maintain the optimal value, facilitating the retrieval of the optimal solution via backtracking if necessary.
After gathering the result, different subchannels are randomly assigned to MTs according to the quantity requirements, and then a final channel allocation matrix $\textbf{X}_{N_m\times{K}}$ for ES $m$ is formed. The detailed process of the algorithm is as follows.

\begin{algorithm}[t]
\small
\caption{Dynamic Programming-based Channel Allocation (DCA)}
\label{algDPCA}
\KwIn{$\mathcal{N}_m$, $K$, $\mathcal{Z}_n$,$v_{n,z}$}
\KwOut{Channel allocation matrix $\mathbf{X}_{N_m\times K}$}

Initialize $dp[n][j] \gets +\infty$ for $n=0,\dots,N_m$, $j=0,\dots,K$, and set $dp[0][0] \gets 0$;

\For{$n=1$ \textbf{to} $N_m$}{
    \For{$j=K$ \textbf{downto} $0$}{
        \ForEach{$z \in \mathcal{Z}_n$}{
            \If{$j \geq z$}{
                $dp[n][j] \gets \min \{ dp[n][j],\ dp[n-1][j - z] + v_{n,z} \}$;
            }
        }
    }
}

$j \gets K$, $\textbf{alloc} \gets \emptyset$; \quad // \textit{Stores selected allocations}

\For{$n=N_m$ \textbf{downto} $1$}{
    \ForEach{$z \in$ reverse($\mathcal{Z}_n$)}{
        \If{$j \geq z$ \textbf{and} $dp[n][j] = dp[n-1][j - z] + v_{n,z}$}{
            Append $(n, z)$ to $\textbf{alloc}$;
            $j \gets j - z$; \textbf{break};
        }
    }
}

Assign subchannels according to $\textbf{alloc}$, forming $\mathbf{X}_{N_m\times K}$;

\Return $\mathbf{X}_{N_m\times K}$;
\end{algorithm}

\subsection{D3QN-based Task Offloading}
\begin{figure} [t]
\centerline{\includegraphics[width=3.6in]{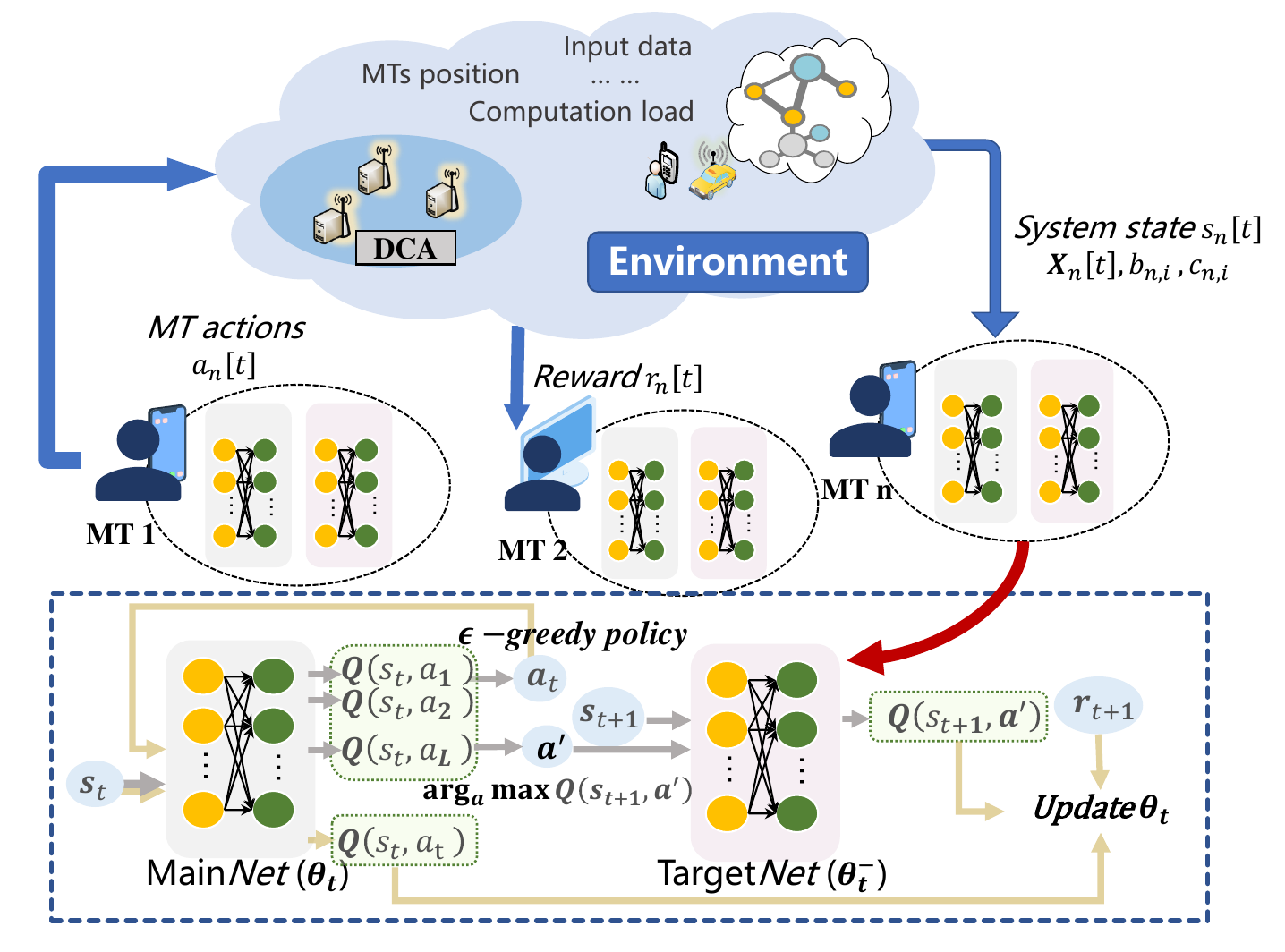}}
\caption{Architecture of D3QN-based Task Offloading.}
\label{fig3D3QNarchitecture}
\end{figure}

As a basic deep reinforcement learning method, DQN addresses the challenge of maintaining the Q table in large state or action spaces in traditional reinforcement learning \cite{2019Human} and has been extensively applied. By utilizing neural networks to approximate the Q value function $\bm{Q}(s_t,a_t;\theta_t)$, it conducts the action selection and learns the strategy that optimizes long-term rewards. 
In essence, a DRL-based method continuously interacts with the environment, adapting in real-time to dynamic conditions, making it well-suited for complex optimization in MEC networks.
Given the large state space and strict latency requirements in this scenario, we adopt D3QN, which is an advanced DQN variant that integrates double DQN and duel DQN or enhanced stability and efficiency. The architecture of the algorithm is displayed in Fig.~\ref{fig3D3QNarchitecture}.
\par

First, the relative elements for D3QN are defined. In time slot $t$, each MT $n$ acts as an agent in the environment, observing the current state $s_n[t]$, choosing an action $a_n[t]$, and then receiving an immediate reward from the environment. Later, the system updates to the next state $s_n[t+1]$. Agents learn the action-choosing policy to maximize the long-term reward by repeatedly interacting with the environment. 
Note that since an MT only performs offloading decisions for one current task indexed by $i$ in each time slot $t$, and the related information of task  $v_{n,i}$ is incorporated into the specific state expression, so $i$ actually appear implicitly in the terms when designing the D3QN-based algorithm.
The detailed principles are introduced in the following part.

\subsubsection{State}
The state of MT $n$ should include the current environment information $\bm{s}_n^{env}[t]$ and the features of pending tasks $\bm{s}_n^{task}[t]$. To be exact, it is the terminal's present position, the amount of data, and the computation, which is as follows
\begin{equation}
\begin{aligned}
           s_n[t]&=\left\{\bm{s}{\rm_n^{env}[t]},\bm{s}{\rm_n^{task}[t]}\right\}=\left\{\textbf{X}{\rm_n[t]}\right\}\cup \left\{b_{n,i},c_{n,i}\right\}.\label{15}
\end{aligned}
\end{equation}
\subsubsection{Action}
The action set for computational task $v_{n,i}$ is
\begin{equation}
    \mathcal{A}={0}\cup\mathcal{M},\label{16}
\end{equation}
where $a_n[t]\in\mathcal{A}$, $a_n[t]=0$ indicates $v_{n,i}$ of MT $n$ at slot $t$ is executed locally while $a_n[t]=m$ means that the task is offloaded to ES $m$ for processing.
\subsubsection{Reward}
The immediate reward that benefits from the proposed subchannel allocation scheme, which is the feedback of action $a_n[t]$ from the environment, is defined as the tanh-scaled cost gain of offloaded processing compared to local execution as follows
\begin{equation}
\begin{aligned}
         r_n[t]&=\tanh(gain_{cost})\\
         &=\tanh(cost_{n,i,0}-cost_{n,i,m^{\prime}}).\label{17}
\end{aligned}
\end{equation}
It aligns with the optimization objective in \eqref{(14)} since the agent is trained to maximize the long-term accumulative reward and thus the negative correlation established between the actual cost and the immediate rewards accomplishes the minimization of the delay-energy weighted cost.

\subsubsection{Algorithm Principles}
Concretely, conventional DQN updates its network parameters as follows
\begin{equation}
    \theta_{t+1}=\theta_t+\eta(y^{DQN}_t-\bm{Q}(s_t,a_t;\theta_t))\nabla_{\theta_{t}}\bm{Q}(s_t,a_t;\theta_t),\label{18}
\end{equation}
while the corresponding updating target is expressed as
\begin{equation}
    y^{DQN}_t=r_{t+1}+\gamma\mathop{\max}_{a^{'}} \bm{Q}(s_{t+1},a';\hat{\theta}_t),\label{19}
\end{equation}
where $\theta_t$ and $\hat{\theta}_t$ is the network parameter of Main\textit{Net} and Target\textit{Net} in \textit{Quasi-static target network} technique, which stabilizes the training \cite{2019Human}. $\eta$ is the learning rate, and $\gamma$ indicates how much the agent attaches importance to future long-term rewards. The above definitions $s_n[t], a_n[t],r_n[t]$ are shortened as $s_t, a_t, r_t$ for the brevity of the expression.

However, the `max' operation and the mode that action selection and valuation in the same network lead to overestimation. Double DQN solves the problem by decoupling them, of which the updating target is rewritten as
\begin{equation}
    y^{DDQN}_t=r_{t+1}+\gamma\bm{Q}(s_{t+1},\mathop{\arg\max}\limits_{a^{\prime}} \bm{Q}(s_{t+1},a^{\prime};\theta_t);\hat{\theta}_t).\label{20}
\end{equation}
It skillfully uses the above two networks to find the action with the maximum Q value in the Main\textit{Net}, while estimating the target Q value in the Target\textit{Net}.

Duel DQN considers that the Q value in certain states may not be related to the action, but will affect the training efficiency. So the estimation of its Q value is divided into two parts: state value function $V(s)$ and the action advantage function $\bm{A}(s,a)$ as follows
\begin{equation}
\begin{aligned}
         \bm{Q}(s,a;&\theta,\alpha,\beta)=V(s;\theta,\alpha)+\\
    &\left(\bm{A}(s,a;\theta,\beta)-\frac{1}{\bm{A}} \sum_{a'}\bm{A}(s,a';\theta,\beta)\right).\label{21}    
\end{aligned}
\end{equation}
It partitions two output branches after the hidden layers of the original DQN, in which $\theta$ stands for the network parameters of the communal part, $\alpha, \beta$ represent the parameters of the two independent fully connected layers for the value function and the action advantage function, respectively. For identifiability, instead of adding the two branches directly, $\bm{A}(s,a;\theta,\beta)$ is decentralized, so the final output could reflect the various roles of $V(s)$ and $\bm{A}(s,a)$.

Bringing the Q value calculation method of Duel DQN in (\ref{21}) into the update mode of Double DQN in (\ref{20}), the D3QN algorithm that deals with discrete action spaces stably and efficiently is obtained. It fits well with the complex environment and real-time decision requirement in our MEC system, while \cite{19} has also proved the strong generalization ability of D3QN to unfamiliar and rapidly changing environments. 
In addition to the mentioned \textit{Quasi-static target network}, $\epsilon$\textit{-greedy policy} and \textit{experience replay} are applied to achieve the balance between exploration and exploitation, break the correlation between samples, ensuring as well as enhancing the training effect. Details of the algorithm are shown below.

\begin{algorithm}
\small
\caption{D3QN-based Task Offloading (DTO)}
\label{algDTOA}
\BlankLine
\textbf{Initialization:} Experience pool $D$ with the capacity of $\mathcal{B}$, parameters of Main\textit{Net} and Target\textit{Net}: $\theta_0$, $\hat{\theta_0}$\;
\While{episode$\leq$ E}{
Reset the system state $s_0$, $\bm{ar}=0$\;
\While{t$\leq$ Total slots}{
Take an action  $a_t$ by $\epsilon$-\textit{greedy policy}\;
Interact with the environment (apply \textbf{DCA})\;
Receive the corresponding reward $r_t$\;
Update the system $s_{t+1}$\;
Store the transition $<s_t,a_t,r_t,s_{t+1}>$ into the experience pool $D$\;
Accumulate rewards $\bm{ar}+=r_t$\;
\If{Samples in $D$ are more than $|\widetilde{\mathcal{B}}|$}{
Select a batch of samples randomly applying \textit{Experience Replay}\;
Perform gradient descent on Main\textit{Net} with parameter $\theta_t$\;
\If{Satisfying the updating frequency}{
Replace parameters $\hat{\theta_t}$ in Target\textit{Net} with $\theta_t$\;
}}
}}
\end{algorithm}

\begin{figure} [!t]
\centerline{\includegraphics[width=3.5in]{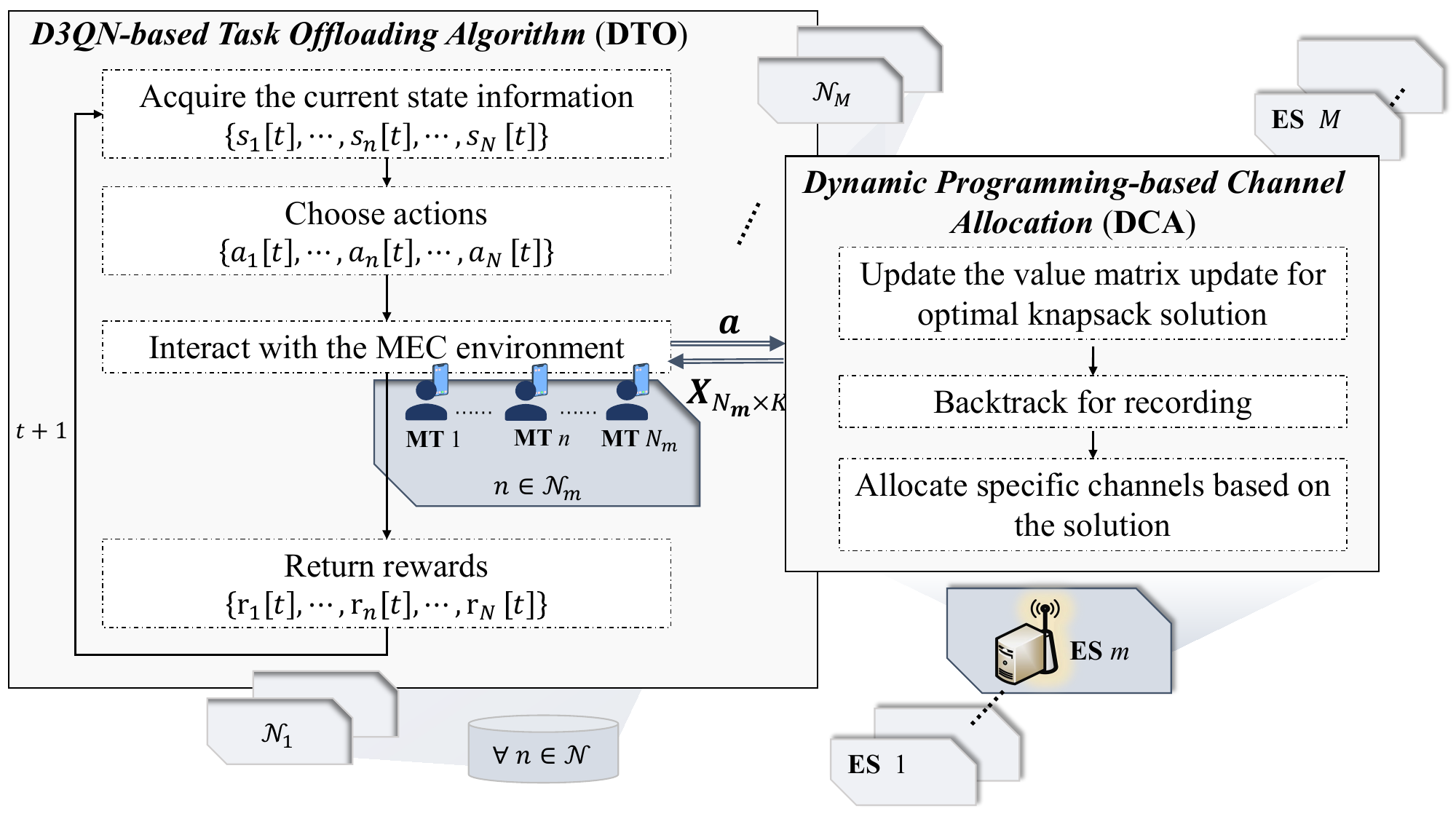}}
	\caption{The structure of the TOICA algorithm.}
	\label{fig4DTOICAstructure}
\end{figure}

The overall process and the deployment manner of the proposed TOICA are shown in Fig.~\ref{fig4DTOICAstructure}.
As an agent, each mobile terminal first observes and obtains the current state information, and then makes action selections under the guidance of the current network parameters and according to the $\epsilon$-greedy strategy. 
Based on the decision results, the task of each MT is processed locally or offloaded in a corresponding manner. 
For MTs accessed to the same edge server, the proposed dynamic programming-based channel allocation is performed, considering their respective transmission requirements.
The returned resource allocation results and current environment status information are used to count corresponding immediate rewards, which are returned to the agent as samples for network training and parameter updates. 
Subsequently, the environment transitions to the next state, and the process repeats for the next time slot.
Within this framework, the complexity of DCA amounts to $\mathcal{O}(N_m(K-N_m+1)(K+1))$, which simplifies to $\mathcal{O}(N_mK(K-N_m))$ when constant terms are removed. The overall complexity of the TOICA algorithm is $\mathcal{O}(ETMN_mK(K-N_m))$, in which $E$ is the number of training phase episodes.
\par

Remarkably, two phases are divided during the following simulations: policy training and decision-making. The former can be accomplished offline to obtain a convergent offloading policy network while the latter is deployed online to make real-time decisions in the dynamic environment.
Through the mutual coupling of the above two algorithms in each round (slot) of training, the TOICA algorithm proposed in this paper is formed.

\section{Performance Evaluation}
To ensure the realism, the parameter settings are derived from established research~\cite{2024comparison,9257019, DAGnew,new1,new2,14new} and practical MEC system specifications. 
Based on typical real-world MEC-enabled applications such as interactive gaming and video analytics, it is assumed that the amount of data for each task is uniformly distributed between 150 and 400 kB, and the computation workload is between 30 and 80~Mcycles.
The computation capacity of mobile terminals is set between 1 and 1.2 GHz, aligning with the processing power of modern mobile devices (e.g., Qualcomm Snapdragon, Apple A-series, and IoT hardware such as Raspberry Pi)\cite{Qualcomm_Snapdragon_865, RaspberryPi4}. 
Their transmission power ranges from 1 to 1.5 W, which falls within the typical uplink transmit power range (20–33 dBm) for mobile and IoT terminals.
The available bandwidth is divided into 8 subchannels ($K=8$), a setting that reflects resource allocation strategies in practical 5G sub-6 GHz deployments \cite{3GPP_TS_38_211}. 
The scenario includes 5 mobile terminals ($N=5$) moving within a 500-meter radius and offloading tasks to 3 edge servers ($M=3$). 
Each terminal executes an application with an average of 15 task nodes ($I_n=15, \forall{n}\in\mathcal{N}$), incorporating task dependency structures to simulate diverse workloads.
The weight factor for delay and energy in the cost function is initially set to $\omega=0.5$.
Additional details of the parameter setting are shown in Table \ref{table1}. 
The simulations are performed on an Intel i5-10210U CPU, the version of Python is 3.8 and the version of torch is 1.8.0. 
To enhance result clarity, the cost in the following simulations is scaled by the $\tanh$ function, which is similar to the scheme of reward setting in (\ref{17}) so that the results can be presented more intuitively.

\begin{table}
\caption{Simulation Parameters}
\setlength{\tabcolsep}{3pt}
\begin{tabular}{|p{90pt}|p{30pt}||p{83pt}|p{26pt}|}
\hline
Parameters& 
Value&
Parameters&
Value\\
\hline
        Total bandwidth $B$ & 50 MHz & Learning rate $\eta$&0.0003\\
        Number of subchannels $K$ & 8  & Discount factor $\gamma$&0.99\\
        Noise power ${\sigma}^{2}$& \scriptsize{-100 dBm} & ${\epsilon}$ decay value & 1.5e-5\\
        MT static power $p_n^{st}$ & 300 mW & minimum ${\epsilon}$ & 0.03 \\
        ES computing capacity $f_m$ & 5 GHz & \scriptsize{Experience pool capacity} $\mathcal{B}$ & 1e6\\
        Degradation factor $\zeta$ & 0.2 & Batch size $\lvert {\tilde{\mathcal{B}}} \rvert$ & 128\\
        Path loss index $\alpha$ & 4 & \scriptsize{Target\textit{Net} update frequency} & 30\\
        \hline
\end{tabular}
\label{table1}
\end{table}

Algorithms for comparison involved in the simulations are:

\begin{itemize}
        \item {\textbf{TOICA-RA}:}
        The task offloading and channel allocation are directed by the proposed TOICA, while the DAG structure is decoupled by a random priority scheme.
        \item \textbf{HRROGA}: Task offloading is performed for the multi-DAG based on the genetic algorithm, while tasks are prioritized by response ratio, as proposed in \cite{2024comparison}.
        \item {\textbf{SEG+DCA}:}
        The single edge greedy offloading algorithm (SEG), as proposed in \cite{9257019}, initially selects the cost-minimized offloading decision for each single and then iteratively adjusts it to achieve the overall best utility. The communication resources are allocated by DCA.
	  \item {\textbf{ON+DCA}:}
	   MT chooses the nearest server to offload its task (offloading nearby, ON), while the corresponding ES's communication resources are allocated by DCA.
\end{itemize}

\begin{figure} [t]
\centerline{\includegraphics[width=\columnwidth]{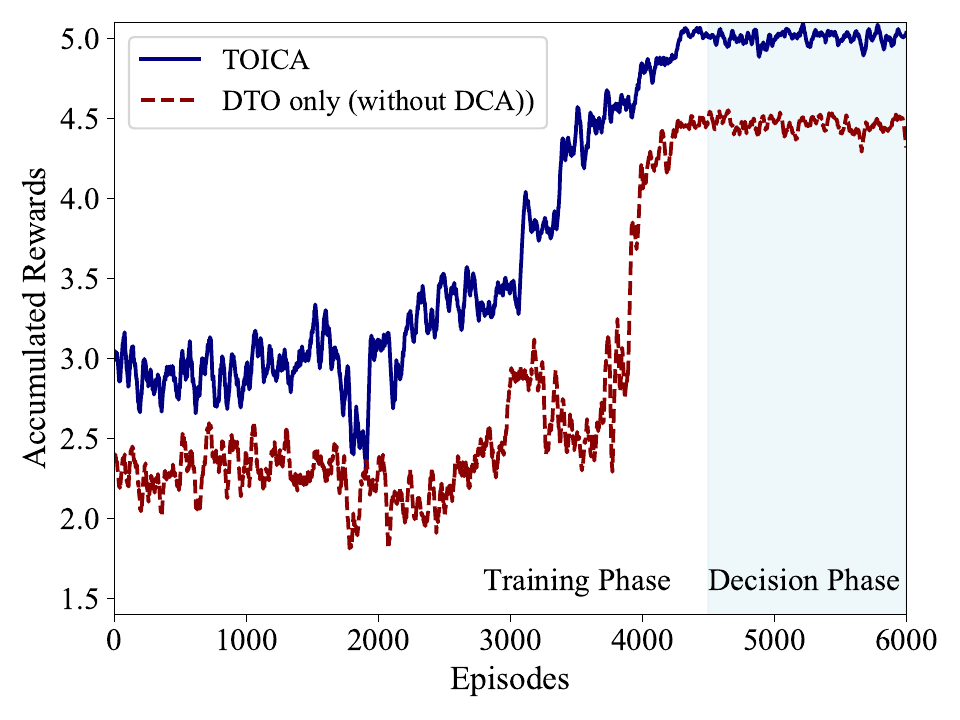}}
	\caption{Progression of cumulative rewards.}
	\label{fig5}
\end{figure}

Fig.~\ref{fig5} displays the convergences of TOICA and DTO Only, in which each episode means the completion of a mobile application for an MT. 
DTO Only indicates applying the proposed DTO to offload tasks while allocating resources in a random manner.
It can be observed that with continuous iterations, the cumulative reward in each episode gradually increases and then stabilizes. 
This reveals that by interacting with the environment, storing experience, training, and updating the network, the terminal has learned an offloading policy that optimizes long-term rewards, verifying the feasibility of the proposed algorithm. Meanwhile, comparing the two curves, it can be observed that the dotted line corresponding to DTO Only has experienced more episodes to be convergent and fluctuates greatly while its cumulative reward is slightly lower than the proposed TOICA. The blindness in resource allocation increases the complexity of learning the optimal strategy, which affects the convergence speed. Also, TOICA considers the amount of data to be transmitted to perform reasonable channel allocation, thereby improving the overall resource utilization of the system while verifying the effectiveness of the DCA algorithm.
Under the current parameters, it takes 551.5 seconds to simulate 1000 episodes, which is an average of approximately 36.8 milliseconds per task. This performance aligns with the response time range (10–100 ms) required for real-time MEC applications such as autonomous driving and real-time video analytics. Given the superior computational power of practical MEC systems, the proposed TOICA method is expected to meet these rapid response criteria in real-world deployments.
\par

\begin{figure} [t]
\centerline{\includegraphics[width=\columnwidth]{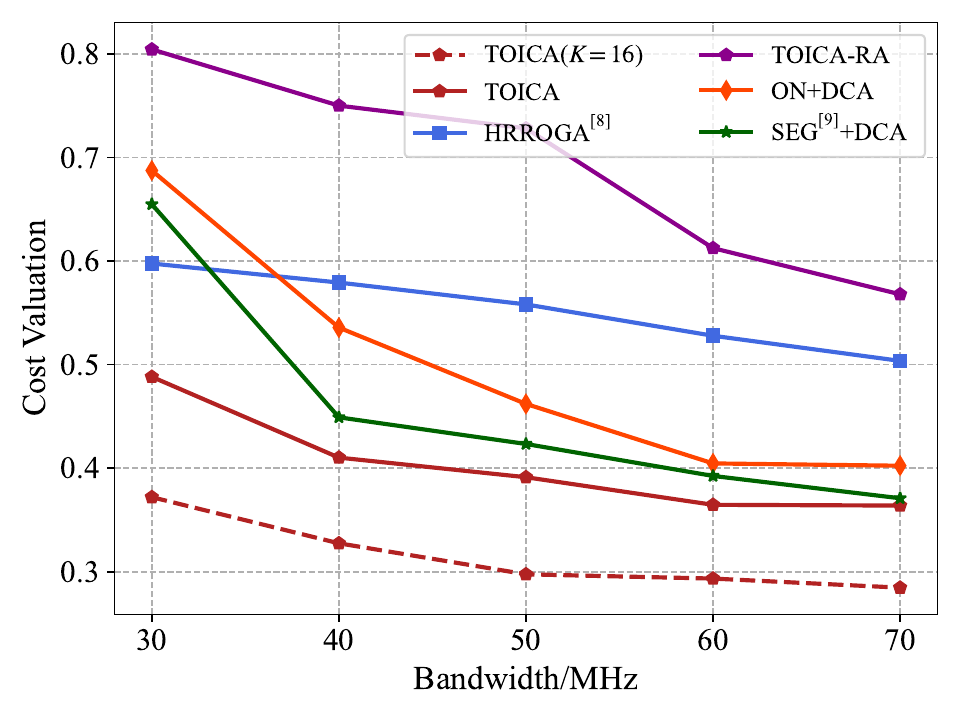}}
	\caption{Cost with varying bandwidth.}
	\label{fig6}
\end{figure}

\begin{figure} [t]
\centerline{\includegraphics[width=\columnwidth]{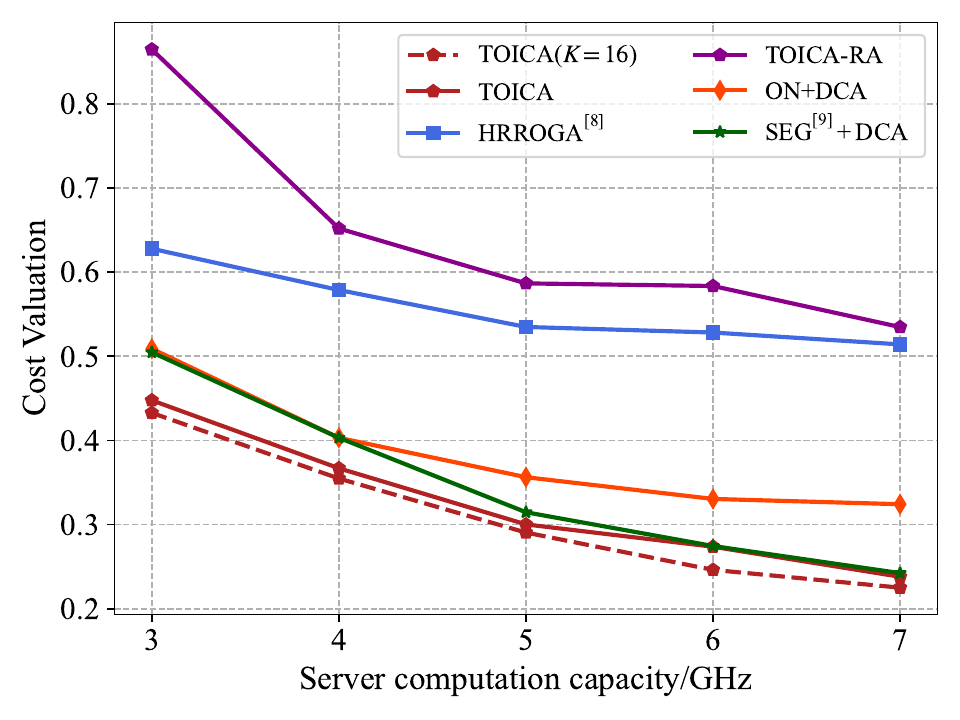}}
	\caption{Cost with varying ES computation capability.}
	\label{fig7}
\end{figure}

Fig.~\ref{fig6} displays the variation of the overall cost with bandwidth $B$ by different algorithms. 
As shown in the figure, the cost tends to decrease as $B$ gets larger because a larger bandwidth indicates a higher data rate and shorter communication time and enhances the system performance. 
What's more, the proposed TOICA conducts the least cost among all the algorithms, while SEG ranks the second lowest, followed by the proximity offloading scheme (ON+DCA).
Because the offloading strategy derived from the Decision Phase of Fig.~\ref{fig5} is more adaptable to the dynamic environment and optimizes the goal more flexibly in a long-term way. 
It is also observed that the performance of both HRROGA and TOICA-RA suffers from their failure to recognize the significance of managing task dependencies in a contextually suitable and goal-consistent way.
The HRROGA decomposes the structured task according to individual subtasks' response ratio, reflecting a less tight relevance to the optimization goal in this work.
TOICA-RA performs poorly due to its disregard for the inherent structural task dependencies, leading to unnecessary efforts in maintaining the application's structural constraints. 
Furthermore, for the same total bandwidth, dividing 16 subchannels has a lower cost than dividing 8 subchannels. The reason may be that increasing the number of subchannels fine-grained the channel allocation process, resulting in greater performance.
\par

Consistent with Fig.~\ref{fig6}, Fig.~\ref{fig7} shows the change of cost with edge service computing capability $f_m$. With its enhancement, the cost valuation shows a downward trend. The improvement of $f_m$ shortens the computing time for processing offloaded tasks, thereby reducing the overall cost. In addition, the proposed TOICA algorithm remains the lowest cost compared with others. The reason is the same as the above analysis, and will not be repeated here.
\par

\begin{figure} [t]
\centerline{\includegraphics[width=\columnwidth]{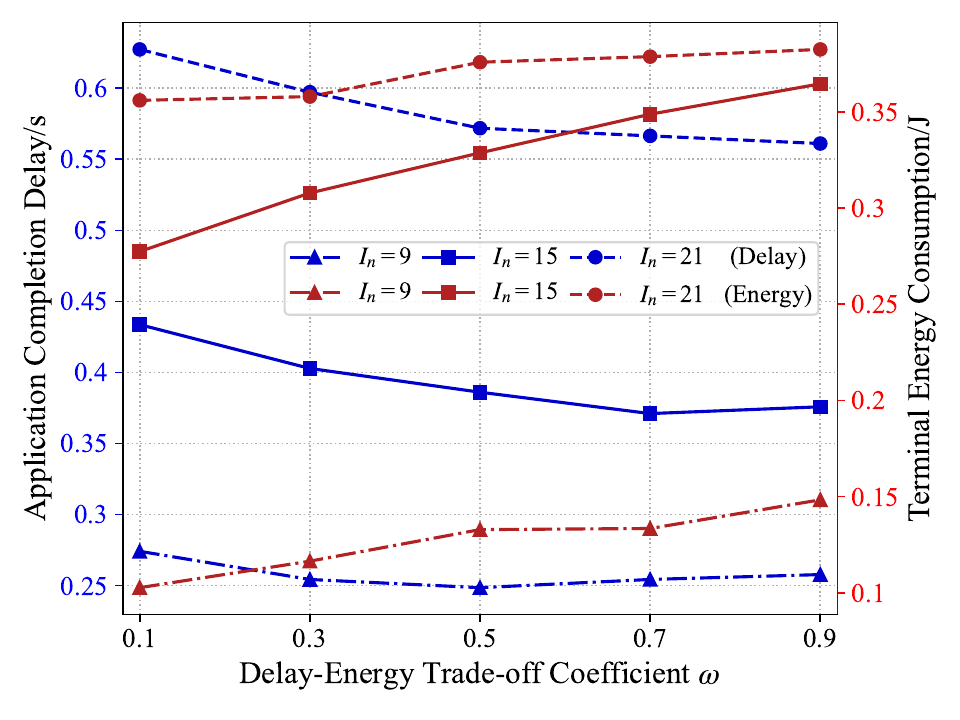}}
	\caption{Delay and energy with varying trade-off coefficient $\omega$ in different numbers of tasks.}
	\label{fig8}
\end{figure}

Fig.~\ref{fig8} first illustrates that how the two specific metrics of application completion delay and energy consumption vary with the trade-off coefficient. It can be seen that with the gradual increase of $\omega$, delay and energy consumption show a decreasing and increasing trend respectively. This fits with our expectation because the larger the $\omega$, the more emphasis is placed on reducing task completion delay while the energy consumption performance deteriorates accordingly. 
This shows that it is allowed to customize the value of $\omega$ based on the specific requirements of different applications. For example, in energy-sensitive scenarios like IoT devices with limited battery capacity, a smaller $\omega$ can be chosen for better energy efficiency. Conversely, for delay-critical applications like VR games, a larger $\omega$ can be selected to minimize task completion delay. 
Fig.~\ref{fig8} also demonstrates the impact of task quantity: as $I_{n}$ increases, both delay and energy consumption rise. 
This correlation is expected as a higher task volume in an application naturally results in more data being transmitted and processed, increasing the delay and energy usage.

\begin{figure} [t]
\centerline{\includegraphics[width=\columnwidth]{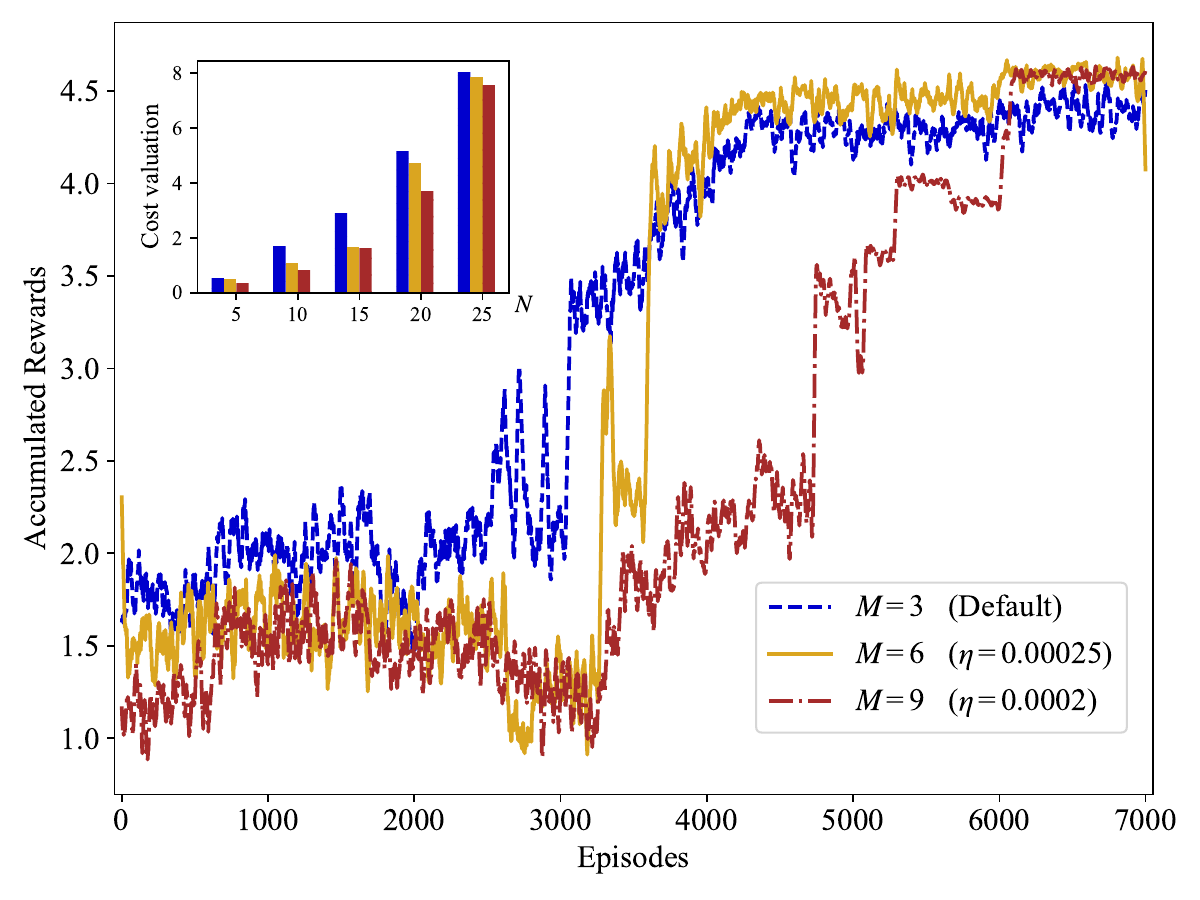}}
	\caption{Progression of cumulative rewards in MEC networks of different scales.}
	\label{fig9}
\end{figure}

Fig.~\ref{fig9} primarily shows the convergence of cumulative rewards for MEC systems with different ESs' numbers. Note that the progress of the curves is the same as that analyzed in Fig.~\ref{fig5}, indicating the feasibility of the proposed TOICA algorithm at different scales.  Given that the number of ESs directly impacts the action spaces and consequently influences the number of neurons in the output layer, we employ distinct hyperparameters for different network structures during the training phase to efficiently approach the optimal strategy. In particular, as the number of ESs increases, network complexity grows, leading us to adopt a smaller learning rate to facilitate stable training.
This careful adjustment of hyperparameters ensures that our TOICA algorithm can effectively handle MEC systems with varying degrees of complexity, providing more potential for adaptability and scalability if combined with advanced parametric adaptive schemes.
Fig.~\ref{fig9} also illustrates the variations in overall cost in scenarios of different scales.  It is evident that when the number of MTs increases under the same number of edge servers, the overall task processing cost of the system experiences a substantial rise due to the limited resources.
Simultaneously, increasing the number of ESs can alleviate the pressure of increased system load to some extent, resulting in successive reductions in overall weighted cost in scenarios with 3, 6, and 9 edge servers, respectively.

\section{Conclusion}
For task offloading in a multi-server multi-user MEC network with spatial-temporal characteristics, this paper presented the task processing model considering the co-channel interference of adjacent cells and the I/O interference of parallel computing in servers. A priority estimation rule based on average cost was designed to convert the DAG structure of dependent tasks into a topological sequence. A delay-energy trade-off cost minimization problem was formulated and solved using a proposed D3QN-based task offloading algorithm integrated multi-cell channel allocation. Simulations verified TOICA's superiority in reducing the delay-energy cost in the spatial-temporal dynamic MEC systems and its scalability for different application scenarios.

Future exploration will focus on the integration of downlink latency considerations, given its emerging relevance in mobile edge computing research, to provide a more comprehensive analysis of MEC systems.
Additionally, deploying DRL-based methods for large-scale and high-density access scenarios will be crucial, necessitating advancements in distributed training, model scalability, and communication efficiency.

{

\bibliographystyle{IEEEtran}

\bibliography{IEEEabrv,TXT_TII-24-0784}

}

\begin{IEEEbiography}[{\includegraphics[width=1in,height=1.25in,clip,keepaspectratio]{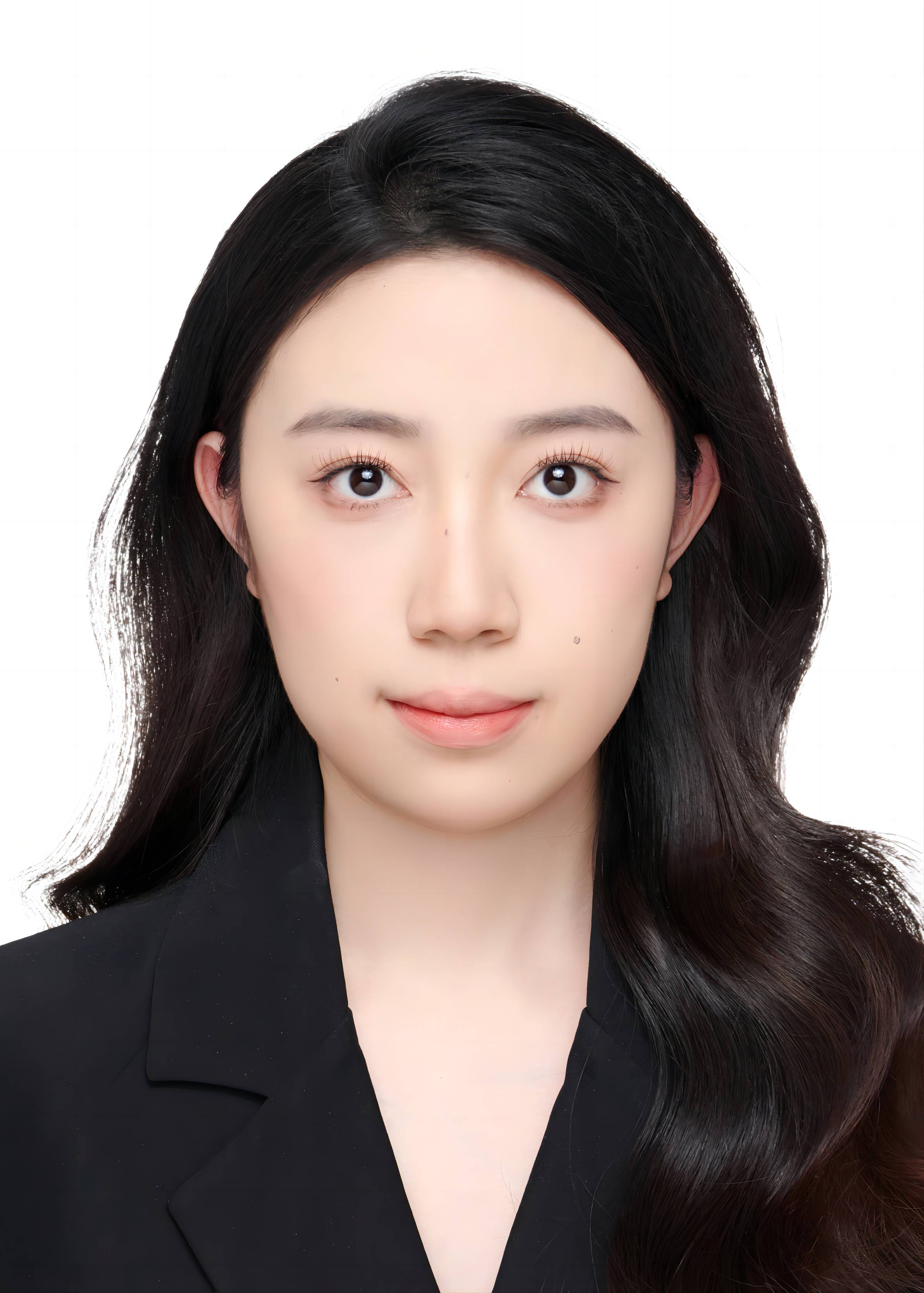}}]{Tianyi Shi} received the B.S. degree in Communication Engineering from Harbin Institute of Technology, Weihai, China, in 2021. She is currently pursuing her Ph.D. degree in Information and Communication Engineering at Beijing University of Posts and Telecommunications. Her research primarily focuses on task offloading and resource allocation in mobile edge computing.
\end{IEEEbiography}

\begin{IEEEbiography}[{\includegraphics[width=1in,height=1.25in,clip,keepaspectratio]{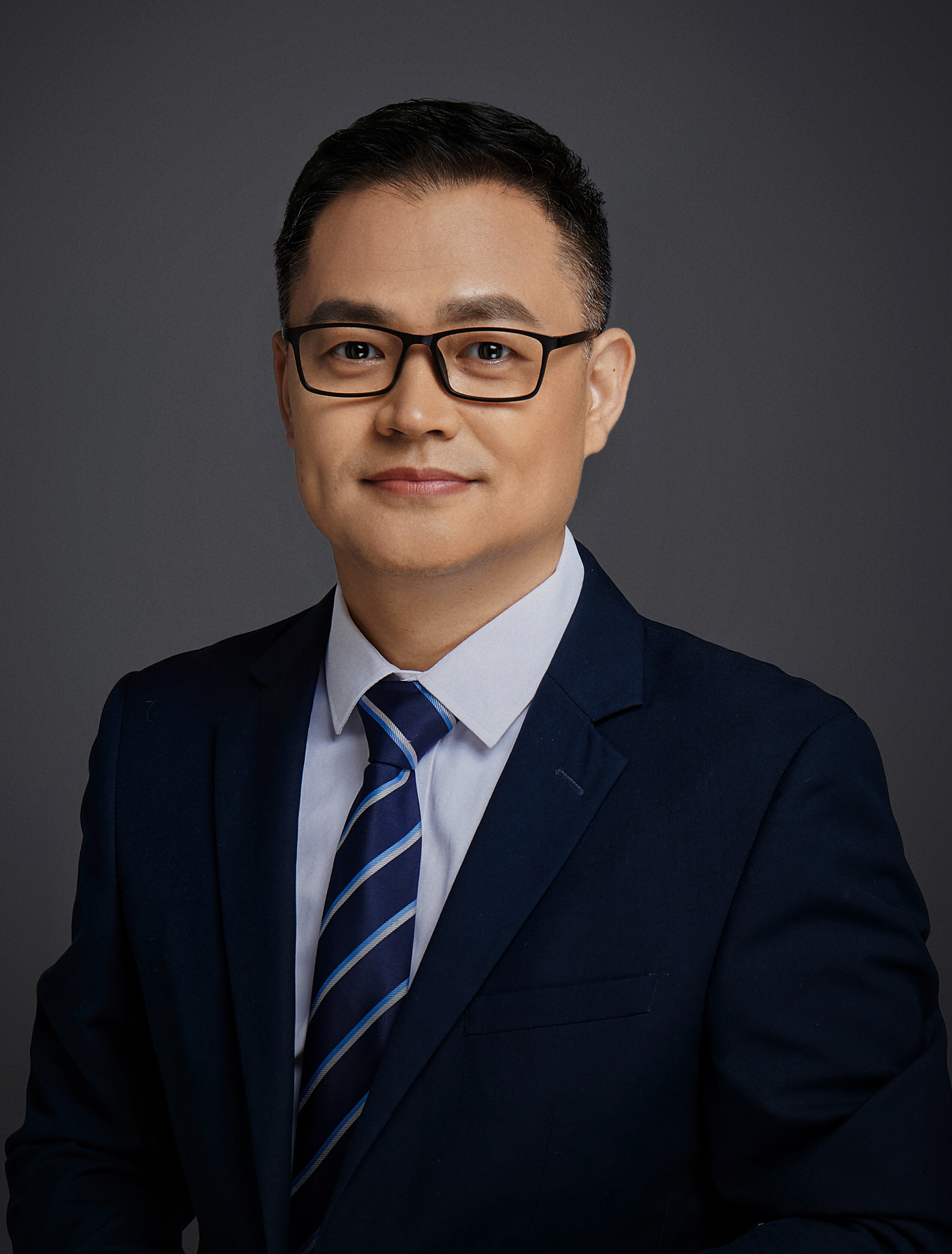}}] {Tiankui Zhang} (M'10-SM'15) received the Ph.D. degree in Information and Communication Engineering and B.S. degree in Communication Engineering from Beijing University of Posts and Telecommunications (BUPT), China, in 2008 and 2003, respectively. Currently, he is a Professor in School of Information and Communication Engineering at BUPT. His research interests include artificial intelligence enabling wireless networks, UAV communications in 5G and beyond networks, intelligent mobile edge computing, signal processing for wireless communications. He had published more than 240 papers including journal papers on IEEE Journal on Selected Areas in Communications, IEEE Transaction on Communications, etc., and conference papers, such as IEEE GLOBECOM and IEEE ICC. He has served as a TPC Member for many IEEE conferences, such as GLOBECOM and PIMRC. He has served as the Technical Program Committee Chair for AiCON 2021. 
\end{IEEEbiography}

\begin{IEEEbiography}
[{\includegraphics[width=1in,height=1.25in,clip,keepaspectratio]{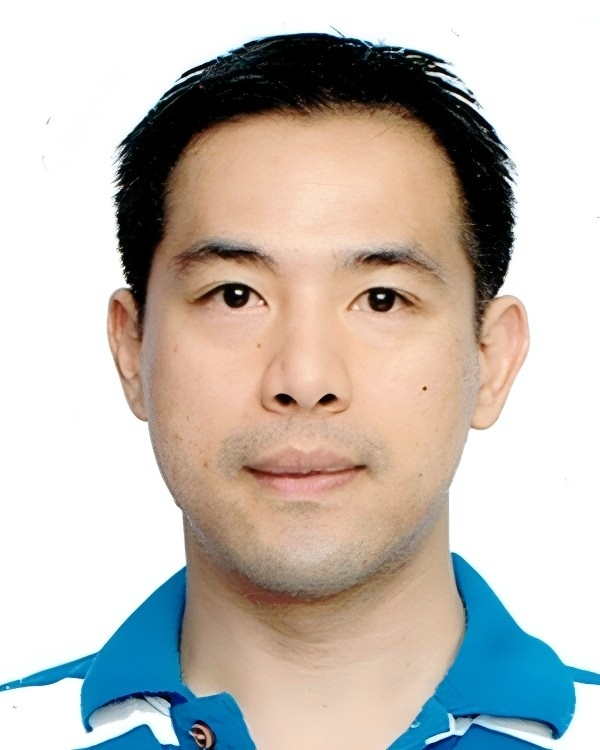}}] {Jonathan Loo} (aka {Kok Keong Loo}) received his M.Sc. degree in Electronics (with Distinction) and the Ph.D. degree in Electronics and Communications from the University of Hertfordshire, UK, in 1998 and 2003, respectively. Between August 2003 and May Jonathan Loo (aka Kok Keong Loo) received his M.Sc. degree in Electronics (with Distinction) and the Ph.D. degree in Electronics and Communications from the University of Hertfordshire, UK, in 1998 and 2003, respectively. Between August 2003 and May 2010, he was a Lecturer in Multimedia Communications at Brunel University, UK. Between June 2010 and May 2017, he was an Associate Professor in Communication Networks at Middlesex University, London, UK. Between June 2017 and June 2024, he was a Chair Professor at University of West London, UK. Since July 2024, he is a Senior Lecturer (Associate Professor) in Electronic Engineering at Queen Mary University of London, UK. His research interests include machine learning for wireless network security, AI-driven IoT system optimization, wireless/mobile networks, network security, wireless communications, and IoT/cyber-physical systems. He has successfully supervised over 20 Ph.D. students as their principal supervisor, and has co-authored more than 380 journal and conference papers in these specialized areas.
\end{IEEEbiography}

\begin{IEEEbiography}
[{\includegraphics[width=1in,height=1.25in,clip,keepaspectratio]{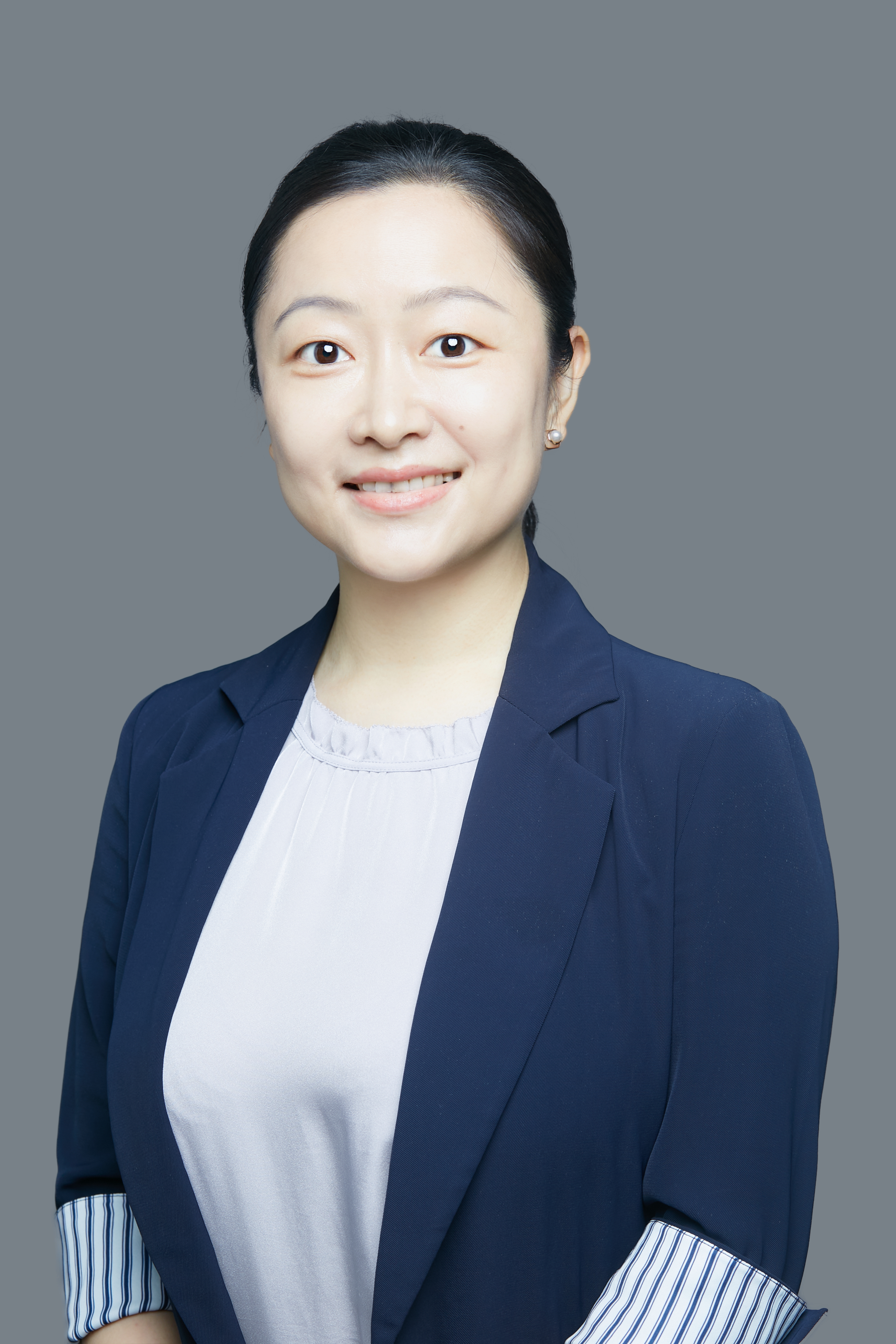}}] {Rong Huang} received the Ph.D. degree in Information and Communication Engineering and B.S. degree in Communication Engineering from Beijing University of Posts and Telecommunications (BUPT), China, in 2013 and 2007, respectively. Currently, she is the chief researcher in mobile communication at the China Unicom Research Institute. Her research interests include intelligent computing, mobile edge computing, AI enabled wireless networks and Industrial IoT. She has been responsible for the technology research in several 5G/6G development Programm of China.
\end{IEEEbiography}

\begin{IEEEbiography}
[{\includegraphics[width=1in,height=1.25in,clip,keepaspectratio]{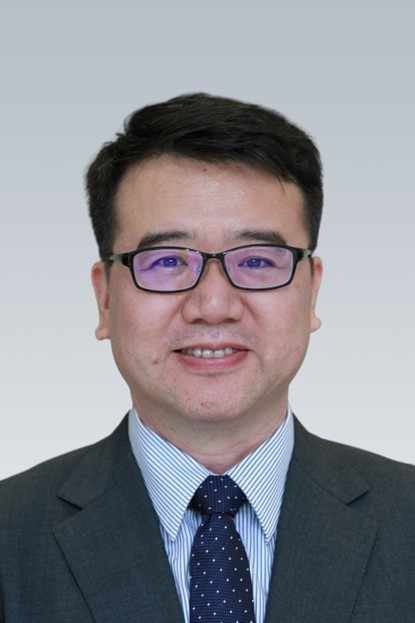}}] {Yapeng Wang} (Member, IEEE, \url{https://fca.mpu.edu.mo/profile/yapengwang}) received a B.Eng. in Telecommunication Engineering and a B.Sc. in Computer and Its Applications from North China Electric Power University, China in 1998, received an M.Sc. in Internet Computing (2002) and a Ph.D. in Electronic Engineering (2007), both from Queen Mary University of London, UK. He joined the Faculty of Applied Sciences, Macao Polytechnic University in 2021 as an associate professor. His current research interests include applied artificial intelligence, wireless communications, automatic speech recognition, nature language processing, medical image analysis, machine learning etc.
\end{IEEEbiography}

\end{document}